\documentclass[aps,prl,twocolumn,groupedaddress]{revtex4-2}

\usepackage{amsmath}
\usepackage{amssymb}
\usepackage{amsfonts}
\usepackage[dvipsnames]{xcolor}
\usepackage{hyperref}
\hypersetup{
	pdftex,
	pdfstartview=FitH,
	bookmarksnumbered=true,
	pagebackref,
	colorlinks,
	breaklinks,
	urlcolor=Maroon,
	linkcolor=Maroon,
	citecolor=Bittersweet
} 
\usepackage{graphicx}
\DeclareGraphicsExtensions{.pdf,.png,.jpg,.mps}
\usepackage{tikz}
\usetikzlibrary{matrix,decorations.pathmorphing,decorations.markings,arrows,positioning,intersections,mindmap,backgrounds}
\usetikzlibrary{positioning,calc} 

\usepackage{booktabs}
\usepackage{physics}
\usepackage{diagbox}

\usepackage{lmodern} 
\usepackage{scalerel,accents,trimclip}
\usepackage{CJK}

\usepackage{shuffle}
\usepackage{cleveref}

\newcommand{\zb}{\bar{z}}

\usepackage{graphicx}

\begin{document}

	\begin{CJK*}{UTF8}{gbsn}

	\title{The Kaluza-Klein AdS Virasoro-Shapiro Amplitude near Flat Space}

	\author{Bo Wang (王波)}
	\email{b\_w@zju.edu.cn}
        \author{Di Wu (吴迪)}
	\email{stevediwu@zju.edu.cn}
	\author{Ellis Ye Yuan (袁野)}
	\email{eyyuan@zju.edu.cn}
	\affiliation{Zhejiang Institute of Modern Physics, School of Physics, Zhejiang University, \\Hangzhou, Zhejiang 310058, China }
	\affiliation{Joint Center for Quanta-to-Cosmos Physics, Zhejiang University,
		\\Hangzhou, Zhejiang 310058, China}

	\date{\today}

	\begin{abstract}
		We bootstrap the first-order correction in the curvature expansion of the Virasoro-Shapiro amplitude in AdS spacetime, for arbitrary Kaluza-Klein charges of external operators. By constructing a universal ansatz based on single-valued multiple polylogarithms as well as an AdS$\times$S formalism, and matching it with the low-lying result, we derive a unified formula in terms of world-sheet integrals. Our result predicts an infinite number of Wilson coefficients that were not available in previous literature.
	\end{abstract}

	\maketitle
	
	\end{CJK*}

    \noindent{\bf Introduction.} The application of analytic bootstrap methods to correlators in holographic conformal field theory models has recently uncovered many hidden structures in these observables in the supergravity limit. In particular, in the prototypical model of $\mathcal{N}=4$ supersymmetric Yang-Mills (SYM), contrary to the na\"ive expectation from complicated bulk interaction vertices, these new developments reveal that these correlators enjoy universal features that are guided by simple principles such as the hidden higher dimensional conformal symmetry \cite{Caron-Huot:2018kta}. This offers a useful probe into the structure of particle interactions in anti-de Sitter (AdS). When taking into consideration the full type IIB superstring dual to SYM, one may wonder whether the above simplicity extends to the entire string scattering \footnote{For recent explorations, see \cite{Drummond:2019odu,Drummond:2020dwr,Aprile:2020luw,Abl:2020dbx,Aprile:2020mus}. }. At least, when we perform curvature expansion on (the Mellin represenations of) the four-point correlator of half-BPS operators, i.e., with respect to $\lambda^{-\frac{1}{2}} = \alpha^{\prime}/R^2$, the leading piece in the genus-0 scattering is well captured by the renowned Virasoro-Shapiro amplitude
    \begin{equation} \label{eq:A0}
    \mathcal{A}^{(0)}(S,T) = -\frac{\Gamma(-S)\Gamma(-T)\Gamma(-U)}{\Gamma(S+1)\Gamma(T+1)\Gamma(U+1)}, 
    \end{equation}
    where $S, T, U$ are the Mandelstam variables, satisfying $S+T+U=0$. It should be ideal that a similar compact formula exists for the complete genus-0 correlator.

    As a promising step towards this goal, recent studies of the stress-tensor correlator $\langle \mathcal{O}_2\mathcal{O}_2\mathcal{O}_2\mathcal{O}_2 \rangle$ as well as its higher Kaluza-Klein (KK) charge contour parts $\langle \mathcal{O}_2\mathcal{O}_2\mathcal{O}_p\mathcal{O}_p \rangle$ show that the corrections in the above curvature expansion can also be expressed as a Riemann sphere integral with insertions \cite{Alday:2023jdk,Alday:2023mvu,Fardelli:2023fyq}. It is then natural to inspect whether these developments can be made systematic for arbitrary KK charge configurations. On the one hand, such an all KK result will definitely help us ease the analysis of the operator spectrum as well as provide new insights on universal structures of the correlators. On the other hand, computation in these general situations is typically not a straightforward task, since existing methods (such as dispersive sum rules \cite{Alday:2022uxp,Alday:2022xwz}) struggle with superconformal block expansion among different KK modes.

    In this work, we resolve the above challenge through a novel AdS$\times$S formalism \cite{Aprile:2020luw}, establishing the world-sheet formula for the first curvature correction to the AdS Virasoro-Shapiro amplitude of arbitrary KK modes. The bootstrap computation leading to this result starts with an ansatz where the world-sheet integrand is built using single-valued multiple polylogarithms (SVMPLs). After constraining the ansatz by crossing symmetry, the remaining degrees of freedom turn out to be uniquely fixed by matching the known $\langle \mathcal{O}_2\mathcal{O}_2\mathcal{O}_p\mathcal{O}_p \rangle$ result \cite{Fardelli:2023fyq}. This circumvents the computational complexity in dispersive sum rules. We validate the universality of our result via low-energy expansions \cite{Drummond:2019odu,Drummond:2020dwr,Aprile:2020luw,Abl:2020dbx,Aprile:2020mus}, where consistency with localization/integrated predictions \cite{Binder:2019jwn,Brown:2023zbr} is confirmed by precise agreement on Wilson coefficients. This world-sheet formula further provides an infinite set of new predictions for the Wilson coefficients.

    \vspace{0.8em}
    
    \noindent{\bf The correlators.} 
    In this paper we consider the supergravitons in $\mathrm{AdS}_5\times\mathrm{S}^5$. On the boundary they are half-BPS operators $\mathcal{O}_{p\geq2}(x,y)$ in $\mathcal{N}=4$ SYM theory, with protected dimension $\Delta=p$ and transforming as $[0,p,0]$ under the $\mathrm{SU}(4)\simeq\mathrm{SO}(6)$ R-symmetry group. Here, $x$ denotes the space-time dependence while $y$ represents the $\mathrm{SO}(6)$ null vector. We can derive the reduced correlator $\mathcal{H}(u,v)$ from four-point correlation function $\langle \mathcal{O}_{p_1} \mathcal{O}_{p_2} \mathcal{O}_{p_3} \mathcal{O}_{p_4}\rangle$ by solving the superconformal Ward identity \cite{Eden:2000bk,Nirschl:2004pa} \footnote{The spacetime cross-ratios are defined as
    \begin{align*}
        u= \frac{x^2_{12}x^2_{34}}{x^2_{13}x^2_{24}} \, ,\quad  v= \frac{x^2_{14}x^2_{23}}{x^2_{13}x^2_{24}} \, ,\quad x^2_{ij} =(x_i-x_j)^2 \, .
    \end{align*}
    }. The reduced Mellin amplitude is defined by
    \begin{align}\label{eq:mellindef}
    \mathcal{H}(u,v;y_{ij})=\int \frac{d s d t}{(2 \pi i)^2} u^{\frac{s+4}{2}} v^{\frac{t-p_{23}}{2}} {M}(s,t;y_{ij}) \Gamma_{p_i} \;,
    \end{align}
    where we abbreviate $p_{ij}=p_i+p_j$ and $y_{ij}=y_i \cdot y_j $. The function $\Gamma_{p_i}$ reads
    \begin{align*}
    \Gamma_{p_i} =&\Gamma\left(\frac{p_{12}-s}{2}\right) \Gamma\left(\frac{p_{34}-s}{2}\right) \Gamma\left(\frac{p_{14}-t}{2}\right) \nonumber \\
    &\Gamma\left(\frac{p_{23}-t}{2}\right)  \Gamma\left(\frac{p_{13}-\tilde{u}}{2}\right) \Gamma\left(\frac{p_{24}-\tilde{u}}{2}\right)\;,
    \end{align*}
    and 
    \begin{align*}
     s + t + \tilde{u}= p_1+p_2+p_3+p_4-4=2\Sigma-4\;.
    \end{align*}
    Note that $s$, $t$ and $\tilde{u}$ are the Mellin analogues of the usual Mandelstam variables for four-point amplitudes in flat space.

    The genus zero contributions of $M(s,t)$ can be decomposed into \cite{Heslop:2022xgp}
    \begin{align}\label{eq:Mgenus0}
    M^{\text{genus-}0}=M^{\text{SG}}+ \lambda^{-\frac{3}{2}}\, M^{3}+ \lambda^{-\frac{5}{2}}\, M^{5}+\cdots
    \end{align}
    where $M^{\text{SG}}$ originates from the supergravity contributions \cite{Rastelli:2016nze} while $M^{i}$ corresponds to higher-derivative stringy effective corrections. The divergences in the asymptotic expansion \cref{eq:Mgenus0} are removed via a Borel transform \cite{Alday:2022xwz}, which resums the perturbative series into a finite amplitude
    \begin{align}\label{eq:boreltransform}
    A(S&,T \, ;y_{ij}) = \lambda ^{\frac{3}{2}} \Gamma(\Sigma-1) \int \frac{d \alpha}{2\pi i}   \frac{e^{\alpha}}{ \alpha^{\Sigma+2}} \nonumber  \\ 
    \times &  M \left(\frac{2\sqrt{\lambda} S }{ \alpha } + \frac{2\Sigma-4}{3} , \frac{2\sqrt{\lambda} T }{ \alpha} + \frac{2\Sigma-4}{3} ;y_{ij} \right)  .
    \end{align}
    This transformation analytically continues the original divergent series into a well-defined amplitude, reorganizing the curvature expansion in $\lambda^{-\frac{1}{2}}$ around flat space,
    \begin{align*}
       {A}^{\text{genus-}0}(S,T\, ; y_{ij})&= \sum_{k=0}^{\infty} \lambda^{-\frac{k}{2}} {A}^{(k)}(S,T \, ; y_{ij}) \, .
    \end{align*}
    The leading term $A^{(0)}$ is proportional to Virasoro-Shapiro amplitude with an extra factor fixed by known supergravity terms on S$^5$ \cite{Rastelli:2016nze,Alday:2023flc}. $A^{(0)}$ exhibits a low-energy expansion whose coefficients lie in the ring of single-valued multiple zeta values (MZVs) \cite{Schlotterer:2018zce,Brown:2018omk,Vanhove:2018elu,Brown:2019wna}. For example, in the $\langle \mathcal{O}_{2}\mathcal{O}_{2}\mathcal{O}_{2}\mathcal{O}_{2}\rangle$ case,
    \begin{align*}
        A^{(0)}_{2222}= \frac{1}{S T U} + 2 \zeta_3 + (S^2+T^2+U^2) \zeta_5 +   S T U \zeta^2_3 + \cdots .
    \end{align*}
    We use $\zeta_n$ to represent the Riemann zeta function. The appearance of single-valued MZVs indicates the structure of world-sheet integrals. Building on this observation one can systematically construct curvature corrections ${A}^{(k)}(S,T)$ by generalizing the integrand on a compact Riemann sphere \cite{Alday:2022uxp,Alday:2022xwz}.
    
    Our goal is to ascertain the first curvature correction $A^{(1)}(S,T)$ for \textit{arbitrary} KK modes. However, for the scattering involving generic KK modes, the dispersive sum rule method necessitates applying superconformal block decomposition and carefully handling the intricate internal $\mathrm{SU}(4)$ symmetry. These difficulties underscore the necessity to enhance the current framework.

    \vspace{0.8em}
    
    \noindent{\bf The AdS$\times$S formalism.} It is convenient to introduce the AdS$\times$S Mellin formalism \cite{Aprile:2020luw} when one deals with the KK levels. This S transform sends $n_{ij}$ to $y_{ij}$ similar to the Mellin transformation which sends $s/t/u$ to $x^2_{ij}$,
    \begin{align}\label{eq:Stranform}
    M(y_{ij})= 
    & \sum_{n_{ij}} \mathcal{M}(n_{ij}) \times \prod_{i<j}\frac{y_{ij}^{n_{ij}}}{\Gamma(n_{ij}+1)} ,
    \end{align}
    where $n_{ij}$ can be understood as the Mellin variables on the sphere S$^5$. To obtain the usual Mellin amplitude, one sums over all integers $n_{ij}$ under the following constraints 
    \begin{align*}
        \sum_{j} n_{ij} = 0\,, \quad n_{ij}=n_{ji}\,, \quad n_{ii} = -(p_i-2) \,.
    \end{align*}
    The gamma functions in the denominator naturally set a cutoff for the range of $n_{ij}$, leading to a finite sum. This formalism simplifies the structures that depend on internal symmetry into problems with finite parameters. This makes it highly effective in diverse scenarios, including the bootstrap process for low-energy expansion \cite{Aprile:2020mus}, computations of 10d effective actions \cite{Abl:2020dbx}, and higher-point calculations \cite{Huang:2024dxr}.

    \begin{figure}[t]
    \centering
    \begin{tikzpicture}[
        node distance = 0.8cm and 2.2cm, 
        every node/.style={draw, rectangle, minimum width=2cm, minimum height=1cm}, 
        solid-arrow/.style={-latex, thick},
        dashed-arrow/.style={-latex, thick, dashed}
    ]

    \node (mathcalM) {$\mathcal{M}(s,t;n_{ij})$};
    \node (M) [right = of mathcalM] {$M(s,t;y_{ij})$};

    \node (mathcalA) [below = of mathcalM] {$\mathcal{A}(S,T;n_{ij})$};
    \node (A) [below = of M] {$A(S,T;y_{ij})$};

    \draw [solid-arrow] (mathcalM) -- (mathcalA);
    \draw [dashed-arrow] (mathcalM) -- (M);
    \draw [solid-arrow] (M) -- (A);
    \draw [dashed-arrow] (mathcalA) -- (A);
    
    \end{tikzpicture}
    \caption{We use solid arrows to represent the Borel transformations \cref{eq:boreltransform} from $\mathcal{M}$ to $\mathcal{A}$ and from $M$ to $A$, and dashed arrows to represent the S transformations \cref{eq:Stranform} from $\mathcal{M}$ to $M$ and from $\mathcal{A}$ to $A$. Since these two transformations act on different variables, we conclude that they are commutative.}
    \label{fig:twoTransform}
    \end{figure}
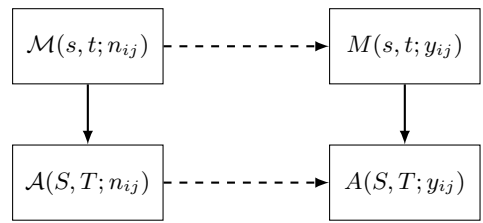

    Within this framework, the tree-level supergravity amplitude for arbitrary KK charges  \cite{Rastelli:2016nze} has a remarkable simplification \cite{Aprile:2020luw},
    \begin{equation}\label{eq:SG}
        \mathcal{M}^{\text{SG}} = -\frac{1}{(\textbf{s}+1)(\textbf{t}+1)(\textbf{u}+1)}  \, ,
    \end{equation}
    here $\textbf{s}$ and $\textbf{t}$ are associated with our notation as
    \begin{align*}
        \textbf{s}=\frac{1}{2}(s-p_{12})+n_{12} \, , \quad  \textbf{t}=\frac{1}{2}(t-p_{14})+n_{14} \, ,
    \end{align*}
    and satisfy the on-shell relation $\textbf{s}+\textbf{t}+\textbf{u}=-4$.
    Regarding the stringy corrections, this formalism can also recast them into a simpler form. For instance, in the case of the $\lambda^{-\frac{3}{2}}$ term, the result for the most general KK charges \cite{Drummond:2019odu,Drummond:2020dwr} can be reorganized as a constant \cite{Aprile:2020luw}
    \begin{align}\label{eq:M13}
        \mathcal{M}^{3} = 2 \,   (\Sigma-1)_3 \, \zeta_3  \, .
    \end{align}

    Having introduced the Borel transform \cref{eq:boreltransform} and the S transform \cref{eq:Stranform}, a useful insight is their commutativity, as illustrated in \cref{fig:twoTransform}. Previous discussions only introduced the Borel transform for $M(s,t;y_{ij})$ \cite{Alday:2022uxp,Alday:2022xwz,Fardelli:2023fyq}, but we suggest considering the transformation from $\mathcal{M}(s,t;n_{ij})$ to $\mathcal{A}(S,T;n_{ij})$ through the Borel transform directly. This framework provides a practical approach for deriving arbitrary KK results while maintaining explicit crossing symmetry.
    
    For instance, the tree-level supergravity contribution to $\mathcal{A}^{(1)}(S,T)$  derived from \cref{eq:SG} yields
    \begin{align}\label{eq:SGA}
        \mathcal{A}^{(1),\text{SG}}&= \frac{2-\Sigma}{6 S^2 T^2 U^2}   \times \left( 3 (n_s S^2 +n_t T^2 +n_u U^2 )\right. \nonumber \\
        & \left. -(n_s+n_t+n_u-1)(S^2+T^2+U^2) \right) \, ,
    \end{align}
    where the crossing symmetric variables are defined as
    \begin{align*}
        n_s=n_{12}+n_{34} \, , \quad n_t=n_{14}+n_{23} \, , \quad 
        n_u=n_{13}+n_{24} \, .
    \end{align*}
    and satisfy the on-shell relation $n_s+n_t+n_u=\Sigma-4$. Furthermore, taking into account the contribution $\mathcal{A}^{(1),i}$ of each string correction $\mathcal{M}^{i}$ under the Borel transform, we can write the full low-energy expansion of $\mathcal{A}^{(1)}(S,T)$ as
    \begin{align*}
       \mathcal{A}^{(1)}(S,T;n_{ij}) = \mathcal{A}^{(1),\text{SG}} + \mathcal{A}^{(1),3} + \mathcal{A}^{(1),5} + \cdots \,.
    \end{align*}
    Notably, deriving $\mathcal{A}^{(1),i}$ from $\mathcal{M}^{i}$ is straightforward. For example, \cref{eq:M13} predicts $\mathcal{A}^{(1),3}=0$. Conversely, reconstructing $\mathcal{M}(s,t)$ from $\mathcal{A}(S,T)$ is uniquely ascertained due to the polynomiality of stingy correction \cite{Aprile:2020mus}. Towards the end of this paper, we will revisit the discussion on Wilson coefficients of $\mathcal{A}^{(1)}$ and provide more precise results.

    The results for the low-lying KK configurations can be reformulated as an integral over the Riemann sphere, suggesting the feasibility of constructing analogous structures in AdS$\times$S space. The supergravity contribution $\mathcal{A}^{(1),\text{SG}}$, as part of the low-energy expansion of $\mathcal{A}^{(1)}$, already delineates a pathway for incorporating $n_{ij}$-dependence into the amplitude. Next, we will present a compact formalism that systematically unifies the treatment of world-sheet integrals and internal symmetry dependencies.
    
    \vspace{0.8em}
    
    \noindent{\bf World-sheet correlator.} Let us construct our ansatz. On the one hand, we choose the single-valued multiple polylogarithms (SVMPLs) as basis, which ensures that the amplitude contains only single-valued MZVs. On the other hand, the coefficients should contain the internal symmetry dependencies. We assume the curvature correction for the world-sheet correlator as
    \begin{equation}
        \mathcal{A}^{(1)}(S,T)= \mathcal{B}^{(1)}(S,T) + \mathcal{B}^{(1)}(S,U)+ \mathcal{B}^{(1)}(U,T) \, ,
    \end{equation}
    which respects the crossing symmetry. $\mathcal{B}^{(1)}(S,T)$ is an integral over the Riemann sphere
    \begin{equation}\label{eq:Riemannsphereint}
        \mathcal{B}^{(1)}(S,T)= \int  d^2z \,|z|^{-2S-2} \,|1-z|^{-2T-2} \mathcal{G} (z,\zb) \, .
    \end{equation}
    In the above $|z|^{-2S-2} \,|1-z|^{-2T-2} $ is the Koba-Nielsen factor, $z$ the complex cross-ratio on the Riemann sphere and the integration measure reads $d^2z=dzd\zb/(-2\pi i)$. 
    
    Our ansatz generalizes the structure of $\langle \mathcal{O}_2\mathcal{O}_2\mathcal{O}_p\mathcal{O}_p \rangle$, with the integrand
    \begin{equation}\label{eq:ansatz}
        \mathcal{G}(z,\zb) = \sum_{i=1}^{4} \mathcal{L}^{s}_{i}\mathcal{R}^s_{i}+\sum_{j=1}^{3} \mathcal{L}^{a}_{j} \mathcal{R}^a_{j} \, .
    \end{equation}
    Here $\mathcal{R}^{a/s}_i(S,T)$ are homogeneous rational functions in the Mandelstam variables $S$, $T$, $U$ and $\mathcal{L}^{s/a}_i$ denote combinations of pure transcendental weight 3 SVMPLs $\mathcal{L}$. Both of them are carefully chosen to ensure consistency with the pole structure and low-energy behavior of the AdS Virasoro-Shapiro amplitude. The symmetric components $\mathcal{L}^{s}_{i}$ and antisymmetric components $\mathcal{L}^{a}_{j}$ under the exchange $z\leftrightarrow 1-z$ \cite{Alday:2023jdk,Alday:2023mvu} are defined as 
    \begin{align}
    \mathcal{L}^{s}_{i}&=\left( \mathcal{L}^{s}_{000}(z),\mathcal{L}^{s}_{001}(z) \,,\mathcal{L}^{s}_{010}(z),\zeta(3) \right)  \, ,\nonumber \\
    \mathcal{L}^{a}_{i}&=\left( \mathcal{L}^{a}_{000}(z),\mathcal{L}^{a}_{001}(z) ,\mathcal{L}^{a}_{010}(z) \right) \, , \nonumber
    \end{align}
    with explicit expressions provided in the Supplementary Material. The prefactors $\mathcal{R}^{a/s}_i(S,T)$ are designed with a universal denominator $U^2$ and enjoyed the general structure
     \begin{equation}
         \mathcal{R}^{a/s}_i(S,T)= \frac{ f^{a/s}_1 S^2 +f^{a/s}_2 \,S \, T +f^{a/s}_3 \, T^2}{(S+T)^2} \, ,
     \end{equation}
    where $f^{a/s}_i$ are linear combinations of 
    \begin{align*}
        \{\Sigma \,n_s,\Sigma \, n_t, \Sigma \,n_u, \Sigma \, , n_s, n_t , n_u \}
    \end{align*}
    with unknown coefficients. This parameterization is motivated by two observations. First, supergravity contributions $\mathcal{A}^{(1),\text{SG}}$ already encode these $n_{ij}$ dependencies. Second, the known $A^{(1)}_{22pp}$ exhibit at most quadratic dependence on $p$, ensuring the sufficiency of the ansatz. Notice that the on-shell relation $n_s+n_t+n_u=\Sigma-4$ reduces the number of basis. This ansatz systematically encapsulates the $\mathrm{SU}(4)$ information for arbitrary KK charges.

     We establish key consistency conditions that constrain the ansatz \cref{eq:ansatz}. The ansatz must satisfy the following fundamental constraints:
    \begin{itemize}
        \item Crossing symmetry: Invariance under the exchange $z \leftrightarrow 1-z$ translates to a symmetry in the Mandelstam variables $S \leftrightarrow T$ and $n_s \leftrightarrow n_t$ enforcing
        \begin{align*}
            \mathcal{B}^{(1)}(S,T ; \, n_s,n_t)=  \mathcal{B}^{(1)}(T,S ; \, n_t,n_s) \, .
        \end{align*}
        The crossing symmetry impose nontrivial relationships among the coefficients of $f^{a/s}_i$, enabling partial determination of their values. 
        \item Single-valuedness: The use of SVMPLs guarantees that the amplitude only contains the single-valued MZVs, a critical requirement for closed-string amplitudes. 
    \end{itemize}
    After imposing the crossing relation, 75 unknown coefficients remain in \cref{eq:ansatz}. All of these coefficients can be uniquely determined by comparing to the known $\langle \mathcal{O}_2\mathcal{O}_2\mathcal{O}_p\mathcal{O}_p \rangle$ result presented in \cite{Fardelli:2023fyq}. The complete bootstrap procedure is presented in the Supplementary Materials. We find that the final result exhibits a striking simplification and offers several insightful physical perspectives.
    
    \vspace{0.8em}
    
    \noindent{\bf Results.} We now present the first curvature correction to the Kaluza-Klein AdS Virasoro-Shapiro amplitude. Although the ansatz was constructed with the full basis of weight-3 SVMPLs, by properly reorganizing the bootstrap result, we observed that it only depends on five linearly independent SVMPLs, 
    \begin{align*}
        \mathcal{L}_{1/4}&= \mathcal{L}^{s/a}_{000}(z) \, ,  \quad  \mathcal{L}_{2}= - \mathcal{L}^{s}_{010}(z) +4 \zeta_{3} \, ,  \\
        \mathcal{L}_{3}&= \mathcal{L}^{s}_{000}(z) - \mathcal{L}^{s}_{001}(z) - \mathcal{L}^{s}_{010}(z) \, , \\
        \mathcal{L}_{5}&= \mathcal{L}^{a}_{000}(z) -\mathcal{L}^{a}_{001}(z) -\mathcal{L}^{a}_{010}(z) \, .
    \end{align*}
    where $\mathcal{L}_{1}$, $\mathcal{L}_{2}$ and $\mathcal{L}_{3}$ are symmetric while $\mathcal{L}_{4}$ and $\mathcal{L}_{5}$ are antisymmetric. This basis automatically eliminates divergences of weight three or lower in the Riemann surface integral \cref{eq:Riemannsphereint}. Specifically, none of these basis elements generate logarithmic singularities of the form $\log^{i}(0)$ for $i\leq3$. This is consistent with the world-sheet toy model proposed by \cite{Alday:2023jdk}, where all such divergences cancel due to the intrinsic properties of closed-string amplitudes. We anticipate analogous cancellations in higher-order corrections (e.g., $\mathcal{A}^{(2)}$), and we leave this analysis for future work. In terms of this basis, the curvature correction takes the elegantly simple form,
    \begin{equation}\label{eq:resultG}
        \mathcal{G}(z,\zb) = (\Sigma-2) \times \sum_{i=1}^{5} \mathcal{L}_{i}\mathcal{R}_{i}  \, , 
    \end{equation}
    where the rational prefactors $\mathcal{R}_{i}(S,T)$ are explicitly
    \begin{align}
        \mathcal{R}_{1} &= \frac{ (n_{s} - n_{t}) (S-T) +2 U }{12 (S+T)} \,  , \quad  \mathcal{R}_{2} = \frac{1}{2} \frac{1}{\Sigma -2} \, , \nonumber \\
        \mathcal{R}_{3} &= \frac{ n_s\,  S+ n_t\, T + n_u\, U }{12 (S+T)} \, , \, \,
        \mathcal{R}_{4} = \frac{n_{s}-n_{t}}{12} \,  , \nonumber \\
        \mathcal{R}_{5} &= \frac{n_{s} S- n_{t} T + (n_{u}+2) (T-S)}{12 (S+T) } \, .
    \end{align}

    This result yields several interesting insights. First, an overall prefactor $\Sigma-2$ appears in the integrand. In fact, this factor emanates from the supergravity contribution \cref{eq:SG}. All basis elements $\mathcal{L}_i$ involve the polar contributions such as $1/S^4$ or $1/T^4$ except $\mathcal{L}_2$. These polar contributions are related to \cref{eq:SG}. This illustrates why $\mathcal{R}_2$ looks different. Second, our ansatz reproduces the supergravity result automatically without explicitly imposing additional conditions. This alignment arises from the hidden 10d conformal symmetry of supergravity \cite{Caron-Huot:2018kta}, which is preserved when extended to AdS$\times$S framework.
    
    \vspace{0.8em}

    \noindent{\bf Low-energy expansion.} 
    To derive the low-energy expansion of $\mathcal{A}^{(1)}$ from the world-sheet correlator, we first need to compute the expansion of $\mathcal{B}^{(1)}(S,T)$ around $S=T=0$. This requires evaluating the integrals over basis
    \begin{equation}
    I_w(S,T)= \int d^2z |z|^{-2S-2}|1-z|^{-2T-2} {\cal L}_w(z)\, .
    \end{equation}
    These integrals were computed in \cite{Alday:2023jdk} using the method developed in \cite{Vanhove:2018elu}, yielding the following result
    \begin{align*}
    I_w= \text{polar}+\sum_{i,j=0} (-S)^i (-T)^j \hspace{-15pt} \sum_{W\in 0^i \shuffle 1^j \shuffle w} \hspace{-15pt} \left( {\cal L}_{0W}(1)-{\cal L}_{1W}(1) \right).
    \end{align*}
    Here the polar term contains poles on $S$ or $T$, arising from logarithmic divergences of the integrand either around $z=0$ for $w=0^n$ or around $z=1$ for $w=1^n$. In our case we have
    \begin{align*}
        \text{polar}(0^n)=-\frac{1}{S^{n+1}}\, , \quad  \text{polar}(1^n)=-\frac{1}{T^{n+1}}\, .
    \end{align*}
    \( \shuffle \) denotes the shuffle product.
    In our case, \( 0^i \shuffle 1^j \shuffle w \) represents all possible ways to merge the sequences \( 0^i \), \( 1^j \), and \( w \), while maintaining their individual orderings. An example of the integral is
    \begin{align*}
        I_{000}=-\frac{1}{S^4}-\left( 8S+6T \right) \zeta_5 -\left( 8ST+6T^2 \right) \zeta_3^2 +\cdots .
    \end{align*}

    Now we would like to systematically analyze the low-energy behavior of the Mellin amplitude $\mathcal{M}$ and its Borel-transformed counterpart $\mathcal{A}$. The behavior of $\mathcal{M}$ under Borel transform is dominated by the power of $s$, $t$ and $\tilde{u}$. Critically, the large-p limit \cite{Aprile:2020luw} imposes an upper-bound on the power of $\textbf{s}$ while the perturbative order we considered constrains the lower-bound of $\textbf{s}$. This dual constraint ensures that only a small part in $\mathcal{M}$ contributes to $\mathcal{A}^{(1)}$, allowing us to categorize contributions as leading or sub-leading based on the degree of $\textbf{s}/\textbf{t}/\textbf{u}$ in $\lambda^{-1}$ expansion. The leading terms in $\lambda^{-\frac{i}{2}} \mathcal{M}^{i}$ correspond to homogeneous symmetric polynomials in $\textbf{s}$, $\textbf{t}$ and $\textbf{u}$ of degree $i-3$. These terms are fully determined by the flat-space Virasoro-Shapiro amplitude $\mathcal{A}^{(0)}$ through the Borel transform, contributing dominantly to $\mathcal{A}^{(0)}$ and  partially to $\mathcal{A}^{(1)}$. Sub-leading terms, however, involve mixed polynomials of lower degree in $\textbf{s}/\textbf{t}/\textbf{u}$ and linear dependence on $n_s/n_t/n_u/\Sigma$,
    \begin{align*}
        (\textbf{s}^j \, n_s + \textbf{t}^j \, n_t + \textbf{u}^j \, n_u) \, \mathcal{S}^k \, ,\quad  \mathcal{S}^{j+k} \, , \quad \Sigma \, \mathcal{S}^{j+k} \, ,
    \end{align*}
    where $j+k=i-4$ and $\mathcal{S}^m$ denote a homogeneous symmetric polynomial in $\textbf{s}$, $\textbf{t}$ and $\textbf{u}$ of degree $m$. Constructing a complete basis for these terms requires systematic enumeration of symmetric invariants, as detailed in \cite{Aprile:2020mus}.

    By plugging in our result \cref{eq:resultG} into the integral, in principle we can predict sub-leading Wilson coefficients in $\mathcal{M}^{i}$ for arbitrary $i$. The first few terms are worked out as

    \begin{align*}
        \mathcal{M}^{5}&= \left( {\sigma}_{2,0} -5 \, {\sigma}_{1,1}  + \cdots\right) \zeta_5 \, , \\
        \mathcal{M}^{6}&= \left(\frac{2}{3} \, {\sigma}_{3,0} -6 \, {\sigma}_{2,1} + (4\Sigma-2)\, {\sigma}_{2,0}   + \cdots\right) \zeta^2_3 \, ,\\
        \mathcal{M}^{7}&= \left( {\sigma}_{4,0} -14 \, {\sigma}_{3,1} +  \left(8\Sigma-\frac{51}{8} \right) \, {\sigma}_{3,0} + \cdots\right) \zeta_7  \, ,\\
        \mathcal{M}^{8}&= \left( \frac{4}{5} \, {\sigma}_{5,0} -16 \, {\sigma}_{4,1} +  \left(8\Sigma-11 \right) \, {\sigma}_{4,0} + \cdots\right) \zeta_3 \zeta_5 \, ,
    \end{align*}
    the ``$\cdots$" represents the contributions from $\mathcal{A}^{(k\geq2)}$ and we use a shorthand 
    \begin{align*}
        {\sigma}_{a,b} =(\Sigma-1)_{a+3} \left( {\bf s}^a \, n_{s}^b + {\bf t}^a \, n_{t}^b +{\bf u}^a \, n_{u}^b \right) \, .
    \end{align*}
    These expansions perfectly align with the analysis above. Our result $\mathcal{M}^{5}$ and $\mathcal{M}^{6}$ precisely agree with previous bootstrap result \cite{Abl:2020dbx,Aprile:2020mus}, supersymmetric localisation \cite{Binder:2019jwn} and integrated constraints \cite{Brown:2023zbr}. These perfect alignments verify the validity of our approach. The sub-leading terms of $\mathcal{M}^{7}$ and $\mathcal{M}^{8}$ are not fixed in previous studies \cite{Abl:2020dbx,Aprile:2020mus}. It will be ideal to validate these results further from different perspectives, i.e., the OPE data \cite{Gromov:2023hzc,Julius:2023hre,Julius:2024ewf} and some special limits \cite{Alday:2023pzu,Alday:2024xpq}.

    \vspace{0.8em}

    \noindent{\bf Discussion.} This work establishes a universal framework for computing curvature corrections to the AdS Virasoro-Shapiro amplitude with arbitrary KK modes. By utilizing the AdS$\times$S formalism and constructing an ansatz based on weight-3 single-valued multiple polylogarithms, we circumvented the complicated superconformal block expansion, which is a long-standing challenge in correlators of non-identical KK modes. The derived amplitude not only reproduces the known results but also predicts an infinite tower of sub-leading Wilson coefficients. These coefficients encode both bulk string corrections and effective higher-dimensional interactions, offering a systematic tool to probe AdS geometry beyond the supergravity regime.

    The prefactor $\Sigma-2$ in \cref{eq:resultG} hints that, an underlying 10-dimensional conformal symmetry inherited from the supergravity limit may persist in the world-sheet framework, which ensures consistency between stringy corrections and the low-energy effective action. A more in-depth exploration of this symmetry in AdS$\times$S space is desirable.
    
    Our results also provide a pathway to extracting CFT data for massive stringy operators, such as the dimensions of single-trace operators. These data probe the fine structure of $\mathcal{N}=4$ SYM theory and can be directly compared with predictions from integrability methods.

    It is interesting to generalize our method to higher-order corrections such as $\mathcal{A}^{(2)}$ \cite{Fa:2025xxx}, where we need single-valued multiple polylogarithms of weight six. The cancellation mechanism of divergences observed for weight-3 terms (e.g., $\log^3(0)$ terms) needs rigorous validation at higher weights. The generalization to other AdS backgrounds \cite{Alday:2024rjs,Chester:2024wnb,Chester:2024esn} or AdS Veneziano amplitude \cite{Alday:2024yax,Alday:2024ksp} could reveal universal features of AdS string amplitudes. Finally, the higher genus expansion encodes the non-planar effects, we can consider world-sheet correlators on genus-one Riemann surface, while the one-loop supergravity terms were provided in \cite{Alday:2017xua,Aprile:2017bgs,Aprile:2017qoy,Alday:2018kkw,Aprile:2019rep,Alday:2019nin,Alday:2021vfb,Huang:2024rxr}. These advancements aim to bridge perturbative string theory with non-perturbative holography, ultimately forging a comprehensive dictionary for quantum gravity in AdS.

    \vspace{0.8em}

	\begin{acknowledgments}
	\noindent{\bf Acknowledgments.}
	    The authors would like to thank Tobias Hansen, Paul Heslop, Zhongjie Huang, Michele Santagata for useful discussions and Tobias Hansen, Konstantinos C. Rigatos for valuable suggestions on the first draft. We thank Xiaoyu Fa for collaborations on related projects. BW, DW and EYY are supported by National Science Foundation of China under Grant No.~12175197 and Grand No.~12347103. The work of BW is also supported by the National Natural Science Foundation of China under Grant No.~124B2095. EYY is also supported by the Fundamental Research Funds for the Chinese Central Universities under Grant No.~226-2022-00216.
	\end{acknowledgments}

	\bibliography{refs}
    
    \widetext
	\begin{center}
		\textbf{\large Supplemental Materials}
	\end{center}
	\appendix
	\section{Single-valued multiple polylogarithms}
		 Multiple Polylogarithms (or MPLs) $L_{w}(z)$ are labelled by a ``word'' $w$ formed by a set of complex variables $(a_1,a_2,\ldots,a_n)$ as ``letters''. They can be recursively defined through iterated integrals \cite{Goncharov:1998kja,Goncharov:2001iea}
		\begin{align*}
			L_{a_1a_2\ldots a_n}(z)=\int_0^z\frac{\mathrm{d}t}{t-a_1}L_{a_2a_3\ldots a_n}(t)\,, 
		\end{align*}
        where the length of the word $w$ is known as  the transcendental weight. In our case, we consider only words composed of letters  $a \in \{0,1\}$. Under this condition, MPLs can also be defined through differential relations
        \begin{equation*}
            \partial_z L_{0w}(z) = \frac{1}{z} L_w(z)\,,~~~~\partial_z L_{1w}(z) = \frac{1}{z-1} L_w(z)\,,
        \end{equation*}
        within special cases,
        \begin{align*}
		    L_{\varnothing}(z)=1\,,\qquad
		    L_{0_n}(z)=\frac{1}{n!}\log^nz\,.
	    \end{align*}
        
        We construct the single-valued multiple polylogarithms (or SVMPLs) following \cite{Brown:2004ugm}, where a map $\text{sv:}\;L \to {\cal L}$ was introduced. Under this map, ${\cal L}_{w}$ is a linear combination of $L_{w'}(z)L_{w''}(\bar z)$ that preserves the weight and is single-valued. This can be conveniently implemented using PolyLogTools \cite{Duhr:2019tlz}. In this package SVMPL is denoted by \texttt{cG}, and we define $\mathcal{L}_{a_1 a_2 \cdots a_n}(z)=\mathtt{cG[a_1,a_2,\cdots,a_n,z]}$. Noted that  SVMPLs are closed under the transformation $z \to 1-z$, we can split them into symmetric and anti-symmetric parts.  Here we present the construction of ${\cal L}^{s/a}_{w}$ that we used as basis
        \begin{align*}
            \mathcal{L}^{s}_{000}(z)&= \mathcal{L}_{000}(z) + \mathcal{L}_{111}(z) \, ,\\[6pt]
            \mathcal{L}^{s}_{001}(z)&= \mathcal{L}_{001}(z) + \mathcal{L}_{011}(z) + \mathcal{L}_{100}(z) + \mathcal{L}_{110}(z) -4 \, \zeta_3 \, , \\[6pt]
            \mathcal{L}^{s}_{010}(z)&= \mathcal{L}_{010}(z) + \mathcal{L}_{101}(z) + 4 \, \zeta_3  \, ,\\[6pt]
            \mathcal{L}^{a}_{000}(z)&= \mathcal{L}_{000}(z) - \mathcal{L}_{111}(z) \, , \\[6pt]
            \mathcal{L}^{a}_{001}(z)&= \mathcal{L}_{001}(z) - \mathcal{L}_{011}(z) + \mathcal{L}_{100}(z) - \mathcal{L}_{110}(z) + 4 \, \zeta_3 \, , \\[6pt]
            \mathcal{L}^{a}_{010}(z)&= \mathcal{L}_{010}(z) - \mathcal{L}_{101}(z) - 4 \, \zeta_3  \, .
        \end{align*}
        Additionally, we also provide the explicit representations of SVMPLs in terms of MPLs for reader's convenience,
        \begin{align*}
            \mathcal{L}_{000}(z)&=L_{0}(z)L_{00}(\zb) +L_{0}(\zb)L_{00}(z) +L_{000}(z)+L_{000}(\zb) \, , \\
            \mathcal{L}_{111}(z)&=L_{1}(z)L_{11}(\zb) +L_{1}(\zb)L_{11}(z) +L_{111}(z)+L_{111}(\zb) \, , \\
            \mathcal{L}_{100}(z)&=L_{1}(z)L_{00}(\zb) +L_{0}(\zb)L_{10}(z) +L_{100}(z)+L_{001}(\zb) \, , \\
            \mathcal{L}_{010}(z)&=L_{0}(z)L_{01}(\zb) +L_{0}(\zb)L_{01}(z) +L_{010}(z)+L_{010}(\zb) \, , \\
            \mathcal{L}_{001}(z)&=L_{0}(z)L_{10}(\zb) +L_{1}(\zb)L_{00}(z) +L_{001}(z)+L_{100}(\zb) \, , \\
            \mathcal{L}_{011}(z)&=L_{0}(z)L_{11}(\zb) +L_{1}(\zb)L_{01}(z) +L_{011}(z)+L_{110}(\zb) \, , \\
            \mathcal{L}_{101}(z)&=L_{1}(z)L_{10}(\zb) +L_{1}(\zb)L_{10}(z) +L_{101}(z)+L_{101}(\zb) \, , \\
            \mathcal{L}_{110}(z)&=L_{1}(z)L_{01}(\zb) +L_{0}(\zb)L_{11}(z) +L_{110}(z)+L_{011}(\zb) \, .
        \end{align*}
        Each $\mathcal{L}(z)$ is a uniform transcendental weight of 3 and guarantees the correct Wilson coefficients along with the appropriate pole structure of the AdS Virasoro-Shapiro amplitude.

        \section{Bootstrap Kaluza-Klein AdS Virasoro-Shapiro amplitude}

        We present two different methods for deriving the AdS Virasoro-Shapiro amplitude: one based on direct comparison of the world-sheet integrand, and the other on matching low-energy expansions. This appendix details the first method, which relies on the explicit result of the correlator $\langle \mathcal{O}_2 \mathcal{O}_2 \mathcal{O}_p \mathcal{O}_p\rangle$. The first curvature correction for $A^{(1)}_{22pp}(S,T)$ is expressed as
        \begin{equation} \label{eq:res22pp}
            A^{(1)}_{22pp}(S,T) = B^{(1)}_1(S,T)+B^{(1)}_1(S,U)+B^{(1)}_1(U,T)+B^{(1)}_2(S,T)+B^{(1)}_2(S,U) \, .
        \end{equation}
        These contributions are defined through world-sheet integrals of the form
        \begin{align*}
            B^{(1)}_i(S,T) = \frac{y_{34}^{p-2}}{\Gamma(p-1)} \int d^2 z |z|^{-2S-2} |1-z|^{-2T-2}G^{(1)}_i(S,T,z) \, ,  \quad i=1,2 \, ,
        \end{align*}
        and functions $B^{(1)}_{1}$ satisfy the following crossing relation
        \begin{align*}
            B^{(1)}_1(U,T) = B^{(1)}_1(T,U) \, , \quad \Longleftrightarrow  \quad G^{(1)}_{1} (S,T,z)=G^{(1)}_{1} (T,S,1-z) \, .
        \end{align*}
        with $y^{p-2}_{34}$ encoding the KK charge dependence. The integrand $G^{(1)}(z,\zb)$ for the world-sheet correlator can be written as
        \begin{equation}
            G^{(1)}_{i}(S,T,z) = \sum_{u=1}^{4} r^{s}_{i,u} \mathcal{L}^s_{u} + \sum_{u=1}^{3} r^{a}_{i,u} \mathcal{L}^a_{u} \, ,
        \end{equation}
        with rational functions given by
        \begin{align*}
        r^s_{1} &= \frac{1}{12} (-p^2,(p-2)p \, ,p^2-2p-6,24) \, , \\
        r^a_{1} &= \frac{p^2(S-T)}{12(S+T)} (-1,1,1) \, , \\
        r^s_{2} &= \frac{p(p-2)}{12(S+T)} (3S, -2S-T, -2S-T, 0) \, ,\\
        r^a_{2} &= \frac{p(p-2)}{12(S+T)} (3S, -2S+T, -2S+T) \, .
        \end{align*}
        By slightly rescaling the basis $\mathcal{L}^{s/a}$, the coefficients differ from the original formulation. Notably, the absence of $B^{(1)}_2(U,T)$ reflects an explicit breaking of crossing symmetry between $n_s$, $n_t$ and $n_u$ in the AdS$\times$S framework. For example, the S-transform imposes $n_t=n_u=0$ while $n_s\neq0$ in \cref{eq:res22pp} when considering the $22pp$ case, thereby reducing the crossing symmetry. 
        
        Substituting our ansatz $\mathcal{G}(z,\zb)$ into S transform and comparing the resulting expression to the $22pp$ case uniquely determines the rational functions $\mathcal{R}^{s/a}_{i}$. For the symmetric part we have
        \begin{align}
            \mathcal{R}^s_{1}&=(\Sigma -2)\frac{ \mu+ (n_{s} - n_{t}) (S-T) +2 U }{12 (S+T)} \, , \nonumber \\
            \mathcal{R}^s_{2}&= \frac{(2-\Sigma) \mu }{12 (S+T)} \, , \hspace{4pt}  \mathcal{R}^s_{3}= \frac{(2-\Sigma) \mu }{12 (S+T)} -\frac{1}{2} \, , 
            \hspace{4pt} \mathcal{R}^s_{4}=2 \, ,
        \label{eq:result}
        \end{align}
        where
        \begin{align*}
            \mu= n_s\,  S+ n_t\, T + n_u\, U \,.
        \end{align*}
        For the anti-symmetric part we have
        \begin{align}
            \mathcal{R}^a_{1}& =\frac{(\Sigma -2) (n_{s} S- n_{t} T + \nu +(n_{t}-n_{s}) U)}{12 (S+T) } \, , \nonumber \\ \mathcal{R}^a_{2}&=\mathcal{R}^a_{3}=\frac{(2-\Sigma ) (n_{s} S- n_{t} T+ \nu)}{12 (S+T)} \, ,
        \end{align}
        where
        \begin{align*}
            \nu=(n_{u}+2) (T-S) \, .
        \end{align*}
        Reorganizing the basis into $\mathcal{L}_i$ yields the five independent functions reported in the main text. 
        
        Finally, we would like to briefly describe an alternative method based on low-energy expansion matching. If sufficient low-energy expansions with specific KK modes are provided as inputs, the integrand can still be uniquely determined. In our case, the integrand of $\langle \mathcal{O}_2 \mathcal{O}_2 \mathcal{O}_p \mathcal{O}_p\rangle$ encodes its low-energy expansion, hence demonstrating the equivalence of these two approaches. We anticipate that the low-energy expansion matching will also remain an effective tool in future studies.
        
\end{document}